\documentclass[article,nojss,shortnames]{jss}\usepackage[]{graphicx}\usepackage[]{color}
\makeatletter
\def\maxwidth{ %
  \ifdim\Gin@nat@width>\linewidth
    \linewidth
  \else
    \Gin@nat@width
  \fi
}
\makeatother

\definecolor{fgcolor}{rgb}{0.345, 0.345, 0.345}

\usepackage{framed}
\makeatletter
 {\par\unskip\endMakeFramed%
 \at@end@of@kframe}
\makeatother

\definecolor{shadecolor}{rgb}{.97, .97, .97}
\definecolor{messagecolor}{rgb}{0, 0, 0}
\definecolor{warningcolor}{rgb}{1, 0, 1}
\definecolor{errorcolor}{rgb}{1, 0, 0}
\newenvironment{knitrout}{}{} 

\usepackage{alltt}

\usepackage{thumbpdf,lmodern}
\usepackage[utf8]{inputenc}

\usepackage{nicefrac}
\usepackage{amsfonts,amsmath,amssymb,amsthm,amstext}

\newcommand{\indep}{\perp \!\!\! \perp}

\newcommand{\trtparm}{\beta}

\newcommand{\PS}{s(\rX)}
\newcommand{\ps}{s(\rx)}

\newcommand{\hatPS}{\mathfrak{s}(\rX)}
\newcommand{\hatps}{\mathfrak{s}(\rx)}

\newcommand{\n}{n}
\newcommand{\N}{N}
\newcommand{\erX}{X}

\newcommand{\etc}{\textit{etc.}}

\usepackage{booktabs}

\newcommand{\rY}{Y}
\newcommand{\rX}{\mX}
\newcommand{\rz}{z}

\newcommand{\rx}{\xvec}

\newcommand{\samX}{\chi}

\newcommand{\ie}{\textit{i.e.,}~}
\newcommand{\eg}{\textit{e.g.,}~}

\newcommand{\Ex}{\mathbb{E}}
\newcommand{\V}{\mathbb{V}}
\newcommand{\RR}{\mathbb{R}}

\newcommand{\eps}{\varepsilon}

\usepackage{dsfont}

 \DeclareMathOperator{\ND}{N}
 
 \DeclareMathOperator{\UD}{U}

\def \xvec {\text{\boldmath$x$}}    \def \mX {\text{\boldmath$X$}}

\def \gammavec        {\text{\boldmath$\gamma$}}

\usepackage{numprint}
\npthousandsep{'}
\npdecimalsign{.}

\usepackage[]{graphicx}
\usepackage[]{color}
\usepackage{xcolor}

\usepackage{alltt}
\IfFileExists{upquote.sty}{\usepackage{upquote}}{}
\usepackage{paralist}

\definecolor{Red}{rgb}{0.5,0,0}
\definecolor{Blue}{rgb}{0,0,0.5}

\usepackage[final]{changes}

\newcommand*{\arxiv}[1]{\href{https://arxiv.org/abs/#1}{\texttt{{arXiv}:~#1}}}

\title{On the relevance of prognostic \\ information for clinical trials:\\
      A theoretical quantification}
\Shorttitle{On the relevance of prognostic information for clinical trials}
\Plaintitle{On the relevance of prognostic information for clinical trials}

\author{Sandra Siegfried\\Universit\"at Z\"urich \And
        Stephen Senn\\University of Sheffield \And
        Torsten Hothorn\\Universit\"at Z\"urich}
\Plainauthor{Siegfried and Senn and Hothorn}

\Keywords{Clinical trials, Covariate adjustment, Machine learning,  
        Prognostic covariates, Sample size reduction}

\Address{
Sandra Siegfried and Torsten Hothorn\\
Institut f\"ur Epidemiologie, Biostatistik und Pr\"avention \\
Universit\"at Z\"urich \\
Hirschengraben 84, CH-8001 Z\"urich, Switzerland \\
\texttt{Torsten.Hothorn@uzh.ch}\\

Stephen Senn\\
School of Health and Related Research\\
University of Sheffield\\
Sheffield, United Kingdom\\
}

\Abstract{

The question of how individual patient data from cohort studies or
historical clinical trials can be leveraged for
designing more powerful, or smaller yet equally powerful, clinical trials becomes increasingly
important in the era of digitalisation.
Today, the traditional statistical analyses approaches may seem questionable
to practitioners in light of ubiquitous historical covariate information.

Several methodological developments aim at incorporating historical
information in the design and analysis of future clinical trials, most
importantly Bayesian information borrowing, propensity score methods,
stratification, and covariate adjustment.
\replaced{
Recently}{
In the context of observational studies}%
, adjusting the analysis with respect
to a prognostic score, which was obtained from some machine learning
procedure applied to historical data, has been suggested and we study the
potential of this approach for randomised clinical trials.

In an idealised situation of a normal outcome in a two-arm trial with 1:1
allocation, we derive a simple sample size reduction formula as a function
of two criteria characterising the prognostic score: (1)~The coefficient of
determination $R^2$ on historical data and (2)~the correlation $\rho$
between the estimated and the true unknown prognostic scores.  While
maintaining the same power, the original total sample size $n$ planned for
the unadjusted analysis reduces to $(1 - R^2 \rho^2) \times n$ in an
adjusted analysis.

Robustness in less ideal situations was assessed empirically.  We conclude that there
is potential for substantially more powerful or smaller trials, but only when prognostic
scores can be accurately estimated.

}
\IfFileExists{upquote.sty}{\usepackage{upquote}}{}
\begin{document}

\section{Introduction}

Randomised controlled trials (RCTs) are the gold standard design for the
estimation of an average treatment effect of some novel intervention.  The
high level of evidence deducible from such a study, however, comes at a high
price: Large sample sizes are often required to demonstrate an anticipated
treatment effect with sufficient power.  This not only renders many RCTs
financially intensive, but also raises ethical considerations.  An important
goal of methodological research is therefore the development of methods
allowing for a substantial reduction of the overall sample size or to
estimate the treatment effect with higher precision from equally large
trials.

In many contexts, individual patient data from large cohort studies or
previously conducted RCTs have been collected with great effort over long
periods of time.  Such data contain valuable information about the course of
a disease under standard of care or even in untreated patient populations.
When planning a novel RCT, the questions ``if'' and ``how'' such prognostic
information can be leveraged to increase precision or to reduce the
necessary future sample size arise naturally.

Many contributions to contemporary RCT methodology can be understood as
attempts to solve this common problem.  Information borrowing, propensity
score matching and adjustment, stratification and covariate adjustment are
the main strands of research concentrating on the ``how'' part of the
question.  We focus on the ``if'' aspect and try to identify conditions
allowing trials to be smaller through incorporation of historical prognostic
information.  In an idealised normal model, we derive a simple relationship
between the strength of prognostic information contained in historical
controls, the quality of a prognostic score capturing this information, and
the reduction in total sample size or gain in precision achievable by
adjusting for such a prognostic score in an RCT.

\added{
The prognostic score, originally formalised by \cite{Hansen_2008}, represents
a baseline ``risk'' in terms of a summary score of observed covariates. 
More specifically, the score quantifies the expected response under control
conditions, estimated from reference data, \eg~historical control data.  The
concept of prognostic scores can thus be utilised to collapse large number of
covariates, and potentially high-dimensional or unstructured information,
in a composite score.  In
clinical practice, prognostic scores aim to provide a tool for risk
stratification, for example for clinical behaviour of a disease
\citep[\eg~in prostate cancer,][]{Kreuz_Otto_Fuessel_2020} or in the
intensive care unit \citep[\eg the APACHE or FOUR scores,][]{Knaus_1991,Wijdicks_2005}.  
Statistical methods relying on prognostic
scores (\ie~disease risk scores for binary outcomes), are widely employed for
observational studies
\citep{Nguyen_Collins_Pellegrini_2020,Aikens_Greaves_Baiocchi_2020,Wyss_2016,Arbogast_Ray_2011,Arbogast_Ray_2009}
and have since also found application in clinical trials, \eg~for
stratification
\citep{Cellini_2019,Hurwitz_2018,Herrera_2020,Saffi_2014} or
covariate adjustment \citep{Schuler_2021,Branders_2021}.  

Prognostic score methods have strong ties to stratification and covariate
adjustment, where, in practice, little is known about the actual extent of
the efficiency gained by stratification \citep{Kernan_Viscoli_Makuch_1999}
or covariate adjustment
\citep{Steingrimsson_Hanley_Rosenblum_2017,Robinson_Jewell_1991}.  Similar
to information borrowing or propensity score matching and adjustment, the
prognostic score dynamically leverages historical information.

In our work we explore this idea in an exemplary setup to quantify the
benefits, ``if'' prognostic information is leveraged in the statistical
analysis. We present a simple and general situation in
Section~\ref{sec:methods}, and contrast conditions determining the potential
benefits when employing this approach in Section~\ref{sec:results}.
}

\deleted{
The setup studied here has little relationship to information borrowing and
propensity score methods, but strong ties exist to stratification and
covariate adjustment.  Information borrowing has been central to a wide
range of proposed Bayesian methods which aim to dynamically borrow
information from historical data in RCTs, such as power priors
\mbox{\citep{Ibrahim_Chen_2000,Ibrahim_Chen_Gwon_Chen_2015,Gravestock_Held_2017}},
hierarchical models and more recently elastic priors
\mbox{\citep{Jiang_Nie_Yuan_2020}}.  Potential benefits (or lack thereof) regarding
frequentist (type~I error rate and power) or Bayesian terms when employing
such an approach have been the subject of recent discussions
\mbox{\citep{Viele_Berry_Neuenschwander_2014,KoppSchneider_2020,
vanRosmalen_Dejardin_vanNorden_2018,Galwey_2017,
Collignon_Schritz_Senn_2020,Schmidli_Gsteiger_Roychoudhury_2014}}.

Although initially developed outside the clinical setting
\mbox{\citep{Rosenbaum_Rubin_1983,Stuart_Rubin_2008}}, propensity score methods
have found application in leveraging historical information in RCTs to
potentially reduce the required sample size, either by augmenting the trial
control group \mbox{\citep{Yuan_Liu_Zhu_2019,Lin_GamaloSiebers_Tiwari_2018}} or by
adjustment aiming to reduce the variance
\mbox{\citep{Williamson_Forbes_White_2014,Zeng_Li_Wang_Li_2020}}.  For a
comprehensive overview of methods to incorporate historical information, see
\mbox{\cite{Lim_Walley_Yuan_2018}}.

In principle, the aim of propensity score and covariate
stratification and adjustment might coincide, the philosophy and
implementation, however, are fundamentally different.  While propensity
scores leverage factors that are predictive of assignment, stratification into
more homogeneous subgroups, or more generally, adjusting for prognostic
information leverage important prognostic factors aiming to achieve more
precise treatment effect estimates.

In practice, little is known about the actual extent of the efficiency
gained by stratification \mbox{\citep{Kernan_Viscoli_Makuch_1999}} or covariate
adjustment \mbox{\citep{Steingrimsson_Hanley_Rosenblum_2017,Robinson_Jewell_1991}},
and choosing appropriate strata or important covariates while keeping
complexity low might be challenging or impossible.  A more theoretical
overview on covariate adjustment in randomised clinical trials is available
in \mbox{\cite{Senn_2011}}.

Our studied setup has very strong connections with the prognostic
score, originally proposed as a measure to leverage baseline covariates in
observational studies \mbox{\citep{Hansen_2008}}.  For RCTs the prognostic
score can be employed in a similar spirit, representing a continuous
baseline ``risk'' estimated from reference data under control conditions,
\eg~historical control data, and extrapolated to both treatment arms.
The concept of prognostic scores is far from novel and a well-known clinical
tool for risk stratification, such as for example the APACHE score
\mbox{\citep{Knaus_1991}} or the FOUR score \mbox{\citep{Wijdicks_2005}}
serving as intensive care unit scoring systems.  Prognostic scores are
further employed to summarise high-dimensional data, compressing, for
example, transcriptome-based data to stratify for clinical behaviour of a
disease \mbox{\citep[\eg~in prostate cancer,][]{Kreuz_Otto_Fuessel_2020}}.

Incorporating a prognostic score has a similar aim as stratification and
covariate adjustment, but obviates the need to choose prognostically
important factors while allowing for incorporation of potentially
high-dimensional information in a simple manner.  Similar to information
borrowing or propensity score matching and adjustment, the prognostic score
dynamically leverages historical information.

In our work we utilise this idea in an exemplary setup to quantify the
benefits, ``if'' prognostic information is leveraged in the statistical
analysis.  We present a simple and general situation in
Section~\ref{sec:methods}, and contrast conditions determining the potential
benefits when employing this approach in Section~\ref{sec:results}.
}

\section{Methods}\label{sec:methods}

We consider a simple two-arm RCT aiming to estimate the effect of a
treatment on some continuous primary outcome $\rY \in \RR$.  In the trial,
patients were randomly assigned to either the treatment, $\rz = 1$, or the
control arm, $\rz = 0$.  For each patient a set of patient characteristics
$\rX \in \samX$ were retrieved at baseline, from potentially
high-dimensional, structured or unstructured information.
The prognostic score is defined in terms of an unknown function $s:
\samX \rightarrow \RR$ collapsing the $k$ baseline covariates in $\rX$.
Assuming the outcome $\rY$ stems from a normal distribution, we study the
following data-generating process~(DGP)
\begin{eqnarray} \label{eqn:dgp}
  \rY = \alpha + \trtparm \rz + \{\pi \PS + \sqrt{\sigma^2 - \pi^2} \eps\}
        \sim \ND(\alpha + \trtparm \rz, \sigma^2),
\end{eqnarray}
where $\alpha$ is the intercept parameter and $\trtparm$ the treatment
effect we wish to estimate.  The unexplained variability $\sigma^2$ is
decomposed into a structured error term,
\begin{eqnarray} \label{eqn:err}
  \{\pi \PS + \sqrt{\sigma^2 - \pi^2} \eps \}& \sim & \ND(0, \sigma^2), \quad
  \rX \indep \eps,
\end{eqnarray}
consisting of a mixture distribution of a prognostic score $\PS \sim \ND(0,
1)$, which follows a standard normal distribution by assumption, and an
independent standard normal residual $\eps \sim \ND(0, 1)$.
The parameter $\pi \in [0, \sigma]$ governs the fraction of variability
explained by the prognostic score $\PS$.

The standard deviation of the residual, $\sqrt{\sigma^2 - \pi^2}$, depends
on $\pi$, such that the variance $\sigma^2$ of the structured error
term~(\ref{eqn:err}) is constant.  For $\pi = 0$, the residual variance is
$\sigma^2$ and the prognostic score does not impact the outcome in any way. 
For $\pi = \sigma$, the prognostic score $\PS$ accounts for the total
variability and the residual variance is zero.  Values of $\pi \in (0,
\sigma)$ indicate DGPs with different signal-to-noise ratios regarding the
prognostic score $\PS$.  Large values of $\pi \PS$ are associated with large
values of the outcome $\rY$, in both the treatment and control groups.

In rare cases, the prognostic score function $\pi s()$ might be known and
$\pi \ps$ can be used as an offset in~(\ref{eqn:dgp}), when $\rX = \rx$ was
observed for patients in the trial.  The standard error of the treatment
parameter estimate, $\hat{\trtparm}$, and thus also the sample size
necessary to demonstrate a certain clinically relevant effect, only depend
on the residual variance $\sigma^2 - \pi^2$ in this case.  Typically,
neither $\pi$ nor the prognostic score $\ps$ are available and need to be
estimated.  Sometimes it is appropriate to assume a linear model $\pi \ps =
\rx^\top \gammavec$, where an adjusted estimate for the treatment effect
$\trtparm$ is computed from simultaneous estimation with $\gammavec$.
Using trial data, the joint estimation of the treatment parameter $\trtparm$
and $\pi \ps$ is much more difficult, inefficient, or even impossible for
high-dimensional (\eg~microarray data) or unstructured
(\eg~clinical notes and reports) covariates $\rX$
\citep{Zhang_Ma_2019}, thus potentially necessitating an independent
sample for the estimation of $\pi \ps$.

We are interested in the setup, where one was able to obtain an estimate,
$\hatps = \widehat{\pi s}(\rx)$, of $\pi \ps$ either from the literature or
from historical control data.  The latter situation received some interest
recently.  Assuming one has access to data from past trials on the same
outcome $\rY$ and covariates $\rX$ for control patients, $\rz = 0$, many
statistical and machine learning procedures, for example random forests,
neural networks, \etc~can be used to estimate the prognostic score function
from the conditional mean $\hatps = \widehat{\pi s}(\rx) = \widehat{\Ex}(\rY
\mid \rX = \rx, \rz = 0) - \hat{\alpha}$.  Models~(\ref{eqn:dgp}) for
historical controls ($\rz = 0$) regressing on $\rX = \rx$ are associated
with an explained variability of $R^2 = 1 - \nicefrac{\sigma^2 -
\pi^2}{\sigma^2} = \nicefrac{\pi^2}{\sigma^2}$.

Instead of studying properties of specific estimators, we make an assumption
about the joint distribution of the estimated and the true prognostic scores
in terms of a correlation coefficient $\rho \in [0, 1]$ for the relevant
situation $\pi > 0$,
\begin{eqnarray}\label{eqn:ps}
  \left(\hatPS, \pi \PS \right) \sim \ND_2
  \left[\left(\begin{array}{c}
    0 \\
    0
  \end{array}\right), \pi^2 \left(\begin{array}{cc}
    1 & \rho \\
    \rho  & 1
  \end{array}\right)\right].
\end{eqnarray}
The setup $\rho = 0$ corresponds to a failed attempt to estimate the
prognostic score on historical data.  For $\rho = 1$, we obtained an oracle
$\hatPS = \pi \PS$, possibly from some very big data-base.  More
realistically, values $\rho \in (0, 1)$ describe how well the prognostic
model $\hatPS$ characterises the prognostic score $\pi \PS$; the
corresponding mean-squared error is
\begin{eqnarray*}
  \Ex[\{\pi \PS - \hatPS\}^2] = 2 \pi^2 (1 - \rho).
\end{eqnarray*}
This setup also captures a potential distribution drift from the historical
to the trial data: Even if $\hatPS$ is a very accurate estimator of the true
prognostic score on the historical data, a considerable lack of fit on the
trial data, and thus a small $\rho$, might be due to a temporal drift in the
prognostic score which applies to trial but not historical patients.  In the
absence of distribution shift, $\rho$ increases with increasing historical
sample size $\mathfrak{n}$.
\added{
For the sake of completeness, we introduce a symbol for the out-of-sample
(OOS) explained variability one would obtain, for example, by cross-validation or an
additional test sample based on the prognostic model fitted to historical data only:
}
\begin{eqnarray*}
R^2_\text{OOS} =
    1 - \frac{\V\{\rY - \hatPS\}}{\V\{\rY\}}
    = (2 \rho - 1) R^2.
\end{eqnarray*}
\added{
The predicted variance reduction for the trial, following \cite{Borm_2007} and also more
recently \cite{Branders_2021} and \cite{Schuler_2021}, would then be $1 -
\widehat{R^2}_\text{OOS}$.}

In our simple setup, it is straightforward to see that one can replace the
unknown prognostic score $\pi \PS$ by $\rho \hatPS$ in~(\ref{eqn:dgp})
without changing the distribution of the outcome,
\begin{eqnarray*}
\rY & = &
      \alpha + \trtparm \rz + \pi \PS + \sqrt{\sigma^2 - \pi^2} \eps \sim
      \ND(\alpha + \trtparm \rz, \sigma^2) \nonumber \\
    & \stackrel{d}{=} & 
      \alpha + \trtparm \rz + \rho \hatPS + \sqrt{\sigma^2 - \pi^2 \rho^2}
      \eps \sim \ND(\alpha + \trtparm \rz, \sigma^2).
\end{eqnarray*}
For the trial patients, this change means that treating $\hatPS \in \RR$ as
a single observable and random covariate with unknown regression coefficient
$\rho$ leads to a reduction of the residual variance from $\sigma^2$ (in a
model $\rY \mid \rz \sim \ND(\alpha + \trtparm \rz, \sigma^2)$ ignoring
prognostic information) to $\sigma^2 - \pi^2 \rho^2$ (in a model $\rY \mid
\rz, \hatps \sim \ND(\alpha + \trtparm \rz + \rho \hatps, \sigma^2 - \pi^2
\rho^2)$ adjusting for prognostic information $\hatps$) whenever $\pi > 0$
and $\rho > 0$.  At the price of estimating one additional parameter $\rho$
in the linear model $\rY \mid \rz, \hatps \sim \ND(\alpha + \trtparm \rz +
\rho \hatps, \sigma^2 - \pi^2 \rho^2)$, one can expect a considerable
reduction of the residual variance, and therefore more powerful tests and
confidence intervals for $\trtparm$, when employing this method of
adjustment.  The fraction
\begin{eqnarray} \label{eqn:frac}
  \frac{\text{residual variance } \rY \mid \rz, \hatps}
 {\text{residual variance } \rY \mid\rz}
 = \frac{\sigma^2 - \pi^2 \rho^2}{\sigma^2} = 1 - R^2 \rho^2
\end{eqnarray}
of the residual variances with and without adjustment for prognostic
information approximately corresponds to the fraction of necessary sample
sizes to demonstrate a specific clinically relevant treatment effect for any
nominal level and power because the sample size of the $t$-test (we ignore
the estimation of the additional parameter $\rho$ here for simplicity but
will elaborate on this issue in the Discussion and an Appendix)
decreases linearly with the residual variance.
Equivalently, for fixed sample size~$n$ the precision of the
treatment effect estimate increases as the residual variance decreases. 
\added{It should be noted that the classical ``design factor'' 
$1 - \widehat{R^2}_\text{OOS}$ \citep{Borm_2007,Branders_2021,Schuler_2021}
is biased in our setup, because
$1 - R^2_\text{OOS} = 1 - (2 \rho - 1) R^2 \neq 1 - \rho^2 R^2$.
This discrepancy will be demonstrated empirically in Section~\ref{sec:results}.
}

In contrast to classical covariate adjustment,
the relationship between $\rX$ and $\rY$ can be highly nonlinear or
unstructured in our studied setup, for example when $\rX$ represents image
data and a complex deep neural network is used to obtain $\hatPS$. 
Still only a single additional parameter $\rho$ has to be estimated in addition to
the treatment effect $\trtparm$ from the present trial data.  The type~I
error for hypothesis tests on $\trtparm$ is maintained, assuming the test
procedure deals with random covariates in an appropriate way, and thus lack
of type~I error control reported for Bayesian borrowing procedures
\citep{KoppSchneider_2020} is avoided here.

The most important question is: When does it actually pay off to
leverage prognostic information by incorporating
prognostic scores $\hatps$ estimated on historical data?  We assess this
question theoretically and empirically for specific values $R^2 =
\nicefrac{\pi^2}{\sigma^2} \in (0, 1)$ and $\rho \in (0, 1)$ in
Section~\ref{sec:theoretic}.  Furthermore, we study the impact of deviations
from the rather strict distributional assumption~(\ref{eqn:ps}) on the
prognostic score and it's estimate in Section~\ref{sec:empeval}.

\section{Results}\label{sec:results}
\subsection{Theoretical result}\label{sec:theoretic}

\begin{figure}[t!]
\centering
\begin{knitrout}
\definecolor{shadecolor}{rgb}{0.969, 0.969, 0.969}\color{fgcolor}

{\centering \includegraphics[width=\maxwidth,trim=0 10 0 0,clip]{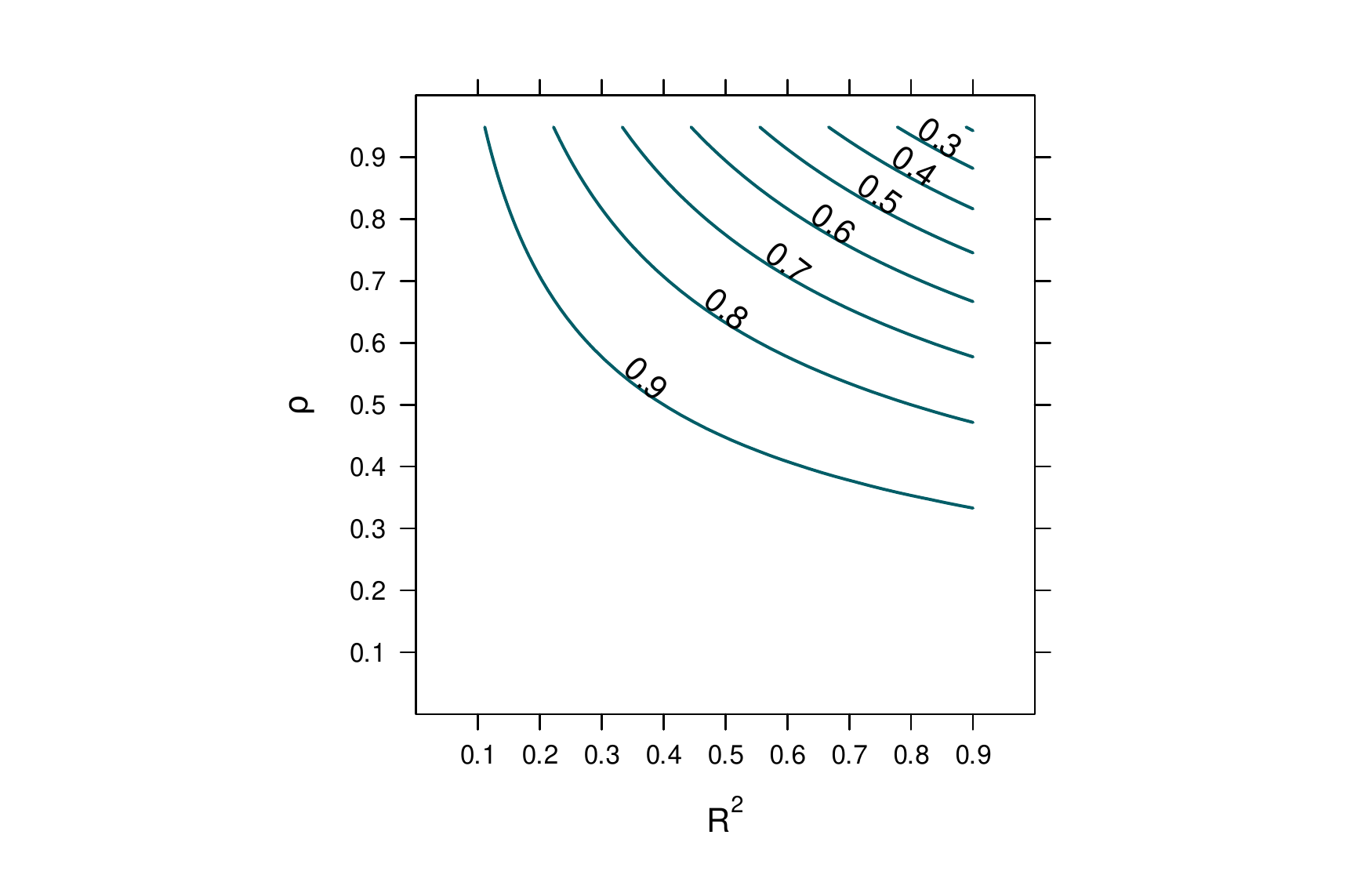} 

}

\end{knitrout}
\caption{Theoretical fraction of residual variances shown for different
values of $R^2 = \nicefrac{\pi^2}{\sigma^2}$ and $\rho$.  The numbers can be
interpreted as the fraction of the sample size required in a trial adjusting
for prognostic information relative to the sample size required for a trial
without such an adjustment.  \label{fig:SIM1}}
\end{figure}

The fraction~(\ref{eqn:frac}) of residual variances with and without
adjustment $1 - R^2 \rho^2$ for values of $R^2 \in (0, 1)$ and $\rho \in (0,
1)$ are presented in Figure~\ref{fig:SIM1}.  The plot can be interpreted as
follows: For a clinical trial powered for the demonstration of a certain
clinically relevant effect in model (\ref{eqn:dgp}) with a specific nominal
level and power, the planned sample size $n$ can be reduced to $(1 - R^2
\rho^2) \times n$ through adjustment for prognostic information.  For
example, with $R^2 = .5$ on a large historical data set resulting in a very
precise estimate $\hatPS$ of $\PS$ with $\rho = .8$, say, only $(1 - .5
\times .8^2) \times 100\% = 68\%$ of the original
sample size $n$ would be required in an adjusted analysis.  Substantial
reductions by more than $20\%$ of the original sample size (\ie $1 - R^2
\rho^2 < .8$) can only be expected for $R^2 > .3$ and rather large values of
$\rho$.  The higher $R^2$, the less precision of the estimate $\hatPS$ is
necessary to achieve the same level of reduction.  For situations with
either small $R^2$ on the historical data and/or small historical sample
sizes $\mathfrak{n}$ resulting in smaller values of $\rho$, expected sample
size reductions of less than $10\%$ (\ie $1 - R^2 \rho^2 > .9$) suggest that
accounting for prognostic information might not be worth the effort.

\subsection{Sensitivity analysis}\label{sec:empeval}
\begin{figure}
\begin{knitrout}
\definecolor{shadecolor}{rgb}{0.969, 0.969, 0.969}\color{fgcolor}

{\centering \includegraphics[width=\maxwidth,trim=0 20 0 0,clip]{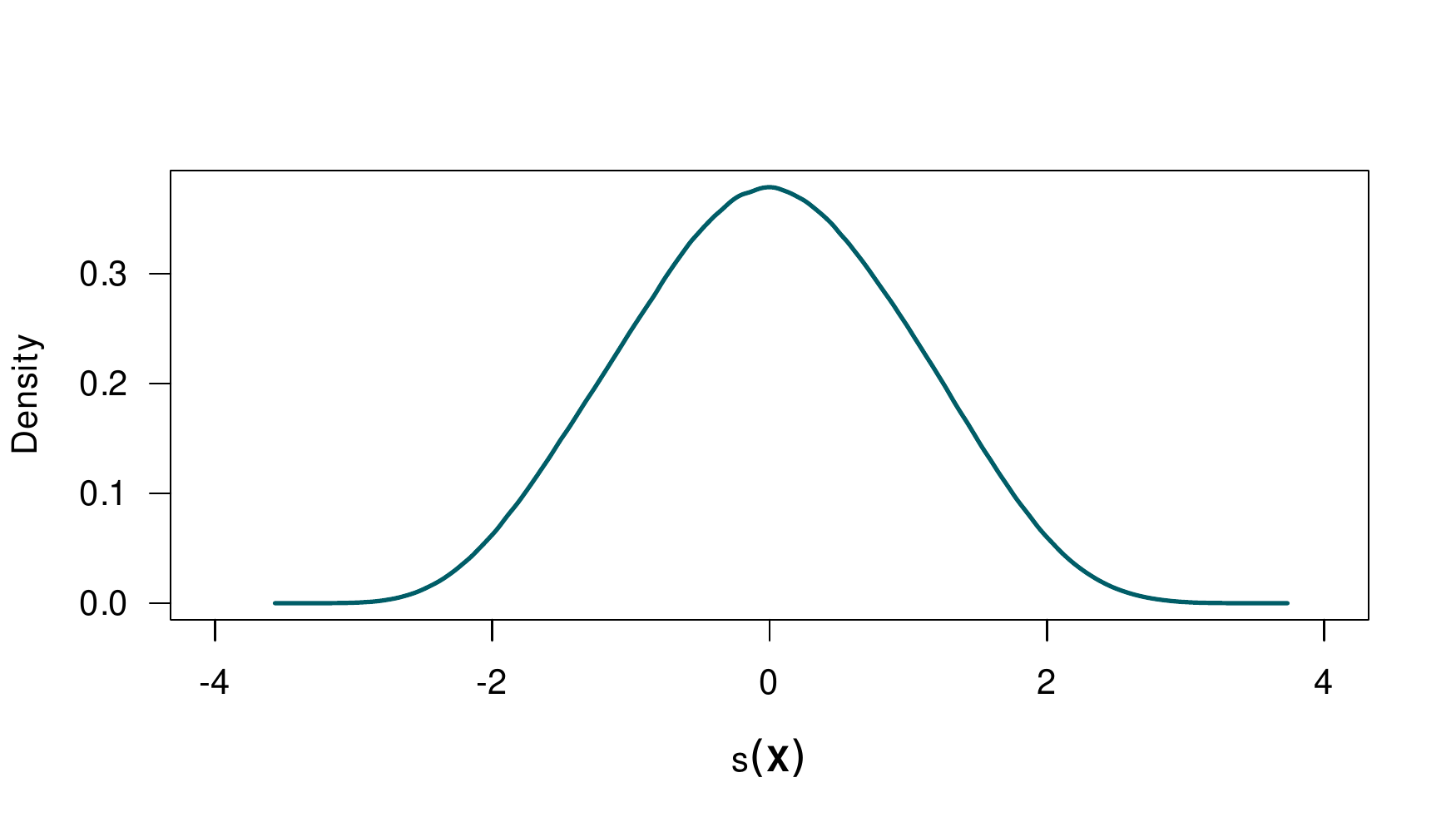} 

}

\end{knitrout}
\caption{Prognostic score.  Illustration of the density function of the
prognostic score $\PS$ drawn from DGP~(\ref{eqn:dgpps}).  The
standardised prognostic score, whilst generated from a non-linear function
of $\rX$ including interaction terms, approximately follows a standard
normal distribution.} \label{fig:pdfps}
\end{figure}

\begin{figure}
\begin{knitrout}
\definecolor{shadecolor}{rgb}{0.969, 0.969, 0.969}\color{fgcolor}

{\centering \includegraphics[width=.8\maxwidth,trim=0 10 0 0,clip]{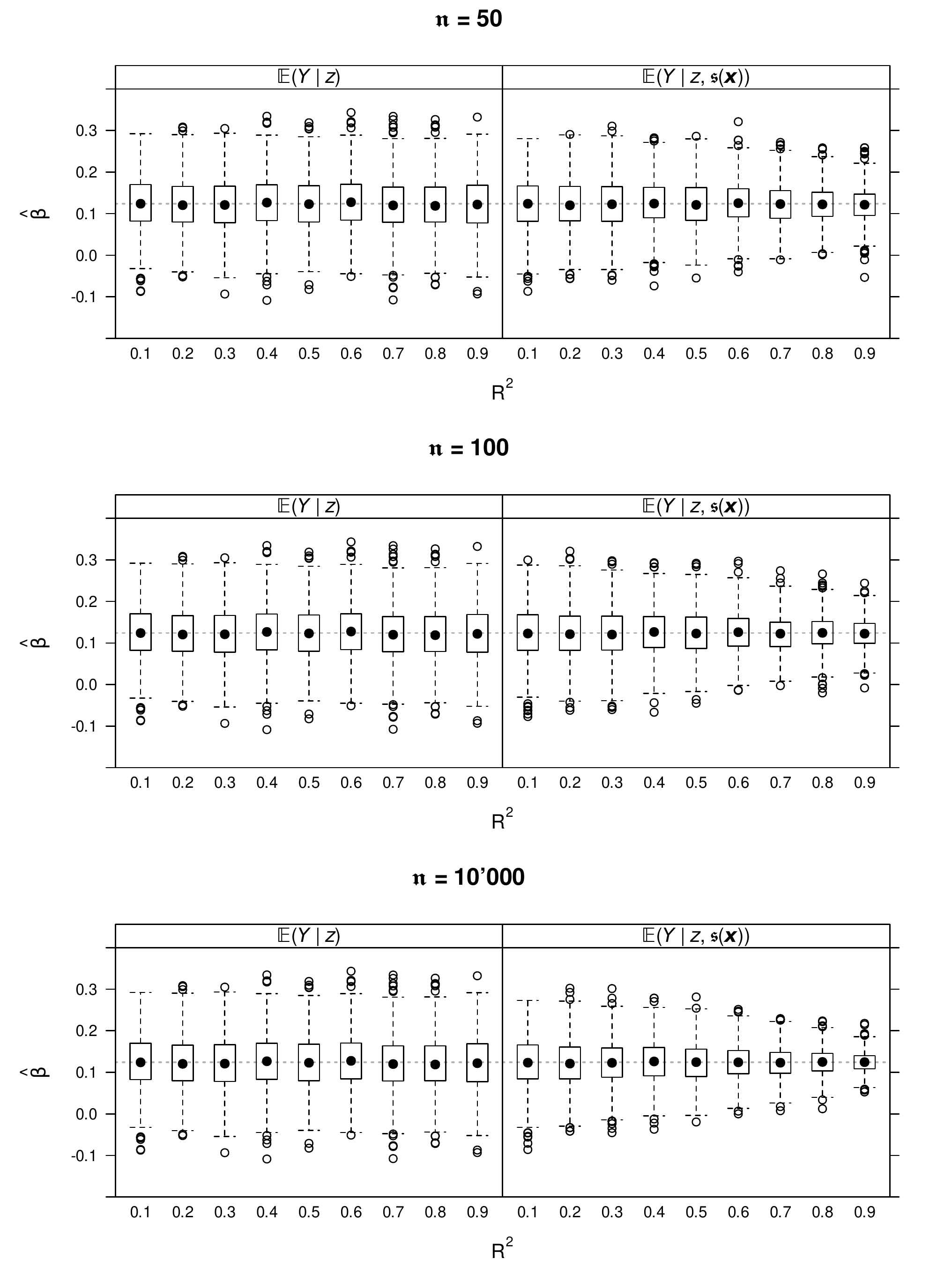} 

}

\end{knitrout}
\caption{Simulated distribution of the treatment effect estimate.  The
treatment effect estimates $\hat{\trtparm}$ from the normal linear model
regressing on the treatment effect ($\Ex(\rY \mid \rz)$; left), and the
model additionally adjusting for the prognostic score estimate ($\Ex(\rY
\mid \rz, \hatps)$; right) are shown for various values of $R^2 =
\nicefrac{\pi^2}{\sigma^2}$ and different sample sizes $\mathfrak{n} = $
\numprint{50}, \numprint{100}, and \numprint{10000}.  The true treatment effect $\trtparm =
0.12$ is indicated by the horizontal line. 
\label{fig:SIM2trt}}
\end{figure}

To study the impact of deviations from the distributional
assumption~(\ref{eqn:ps}), we contrasted the above presented results with a
more complex DGP.  For the prognostic score, we employed the process
\begin{eqnarray} \label{eqn:dgpps}
\PS = 10 \sin(\pi \erX_1 \erX_2) + 20 (\erX_3 - 0.5)^2 + 10 \erX_4 + 5 \erX_5 + \epsilon,
\end{eqnarray} 
arising from Friedman' regression equation~1 \citep{Friedman_1991} with
$\rX \sim \UD(0, 1)^{10}$ and $\epsilon \sim \ND(0, 1)$.
The marginal density of $\PS$ is shown in Figure~\ref{fig:pdfps}.
We simulated historical control data $(\rz = 0)$ of varying sample size
$\mathfrak{n} =$ \numprint{50}, \numprint{100}, and \numprint{10000} as well as trial data with sample
size $n =$ \numprint{1000} from DGP~(\ref{eqn:dgp}) with $\sigma^2 =
1$ for different values of $\pi \in (0, 1)$ and repeated the experiment
\numprint{1000}~times.  We estimated the prognostic model $\hatPS$ from
the simulated historical control data with a random forest and fitted a
normal linear regression model for the treatment effect to the trial data
and a model additionally adjusting for the prognostic score estimate
$\hatps$.
\begin{figure}
\begin{knitrout}
\definecolor{shadecolor}{rgb}{0.969, 0.969, 0.969}\color{fgcolor}

{\centering \includegraphics[width=.8\maxwidth,trim=0 10 0 0,clip]{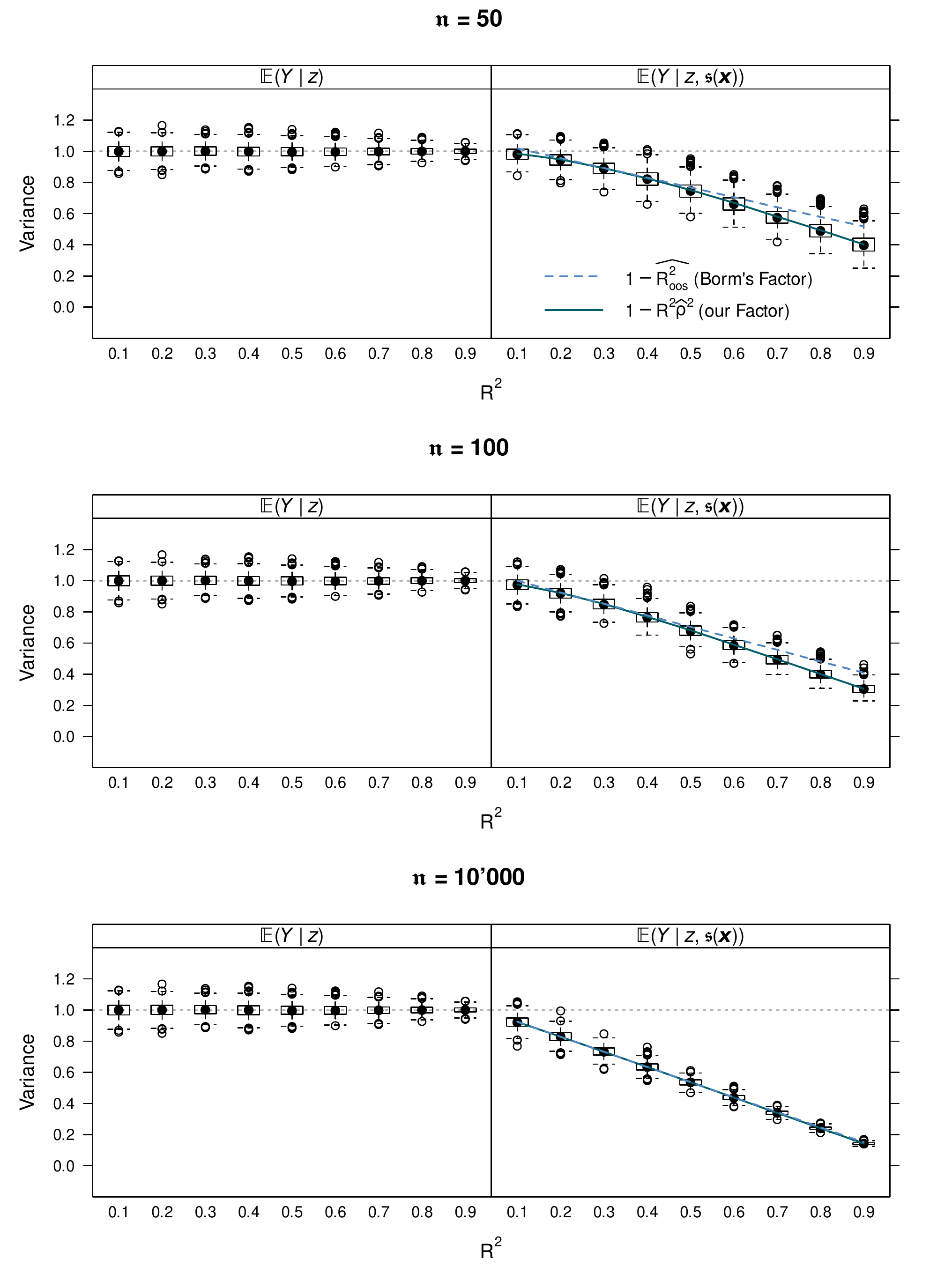} 

}

\end{knitrout}
\caption{Simulated fraction of residual variances in a model with prognostic
score as defined in~(\ref{eqn:dgpps}).  The fractions are shown for the
normal linear model regressing on the treatment effect ($\Ex(\rY \mid \rz)$;
left), and the model additionally adjusting for the prognostic score
estimate ($\Ex(\rY \mid \rz, \hatps)$; right) for various values of $R^2 =
\nicefrac{\pi^2}{\sigma^2}$ and different sample sizes $\mathfrak{n} =$
\numprint{50}, \numprint{100}, and \numprint{10000}. The green line depicts the theoretical fraction $1 - R^2
\hat{\rho}^2$, with the precision of the random forest $\hat{\rho}$
estimated from the data. \added{The variance reduction predicted by the 
``design factor'' \protect{\citep{Borm_2007}} is shown as blue dashed line}.  \label{fig:SIM2var}}
\end{figure}
The results in Figure~\ref{fig:SIM2var} convey similar findings as obtained
theoretically.  The residual variance when adjusting for the prognostic
score estimated from historical data decreases with higher $R^2$, which
translates into higher precision of the treatment effect estimates
$\hat\trtparm$ (Figure~\ref{fig:SIM2trt}).

\subsection{Comparison of predicted and empirical variance reduction}

\added{
We further compared the variance reduction achieved by prognostic score
adjustment as predicted by the ``design factor'' $1 -
\widehat{R^2}_\text{OOS}$ \citep{Borm_2007}, using the estimated
$\widehat{R^2}_\text{OOS}$ from the prognostic random forest model on historical data, to
the empirical variance reduction $1 - \rho^2 R^2$ in our setup.
For the data generating process in Section~\ref{sec:empeval}, 
random forests' $\widehat{R^2}_\text{OOS}$ was estimated
using a large evaluation data set (out-of-sample).  The true $R^2$ was calculated
using $\pi^2 / \sigma^2$.
}


%

\added{
The lines in Figure~\ref{fig:SIM2var} contrast the variance reduction
predicted by the ``design factor'' $1 - \widehat{R^2}_\text{OOS}$ 
\citep{Borm_2007} and $1 - \rho^2 R^2$ (Fraction~\ref{eqn:frac}) 
with the variance reduction achieved empirically (boxplots).  The
latter variance reduction fits the empirical results very closely, whereas
the ``design factor'' is biased and underestimates the actual observed
variance reduction.}

\subsection{Illustration}

A recent study by \cite{Goemans_2020} reported on the development of such a
prognostic score for timed 4-stair climb in Duchenne muscular
dystrophy patients and discussed its potential benefits in terms of design
and analysis of future trials.  The explained variability ($R^2_\text{OOS}$) in the
prognostic model was described to be maximally~$36\%$, which
\added{according to the ``design factor'' would allow for a variance reduction
to $64\%$ of the unadjusted analysis when employing 
prognostic score adjustment. Our derivation, however, indicate that this factor
might underestimate the empirical reduction, which, in practice, is difficult
to quantify, because $R^2$ and $\rho$ are unknown.
}
\deleted{
reportedly would allow for a sample size reduction of
approximately~$40\%$ in an RCT employing prognostic
score adjustment.  Based on our derivations~(see Equation~\ref{eqn:frac})
this reduction would only be attainable in complete absence of
distributional shift, \ie~$\rho = 1$.  However, for trial data deviating
from the training data and thus smaller values of $\rho$ the potential
reduction of the required sample size of equally powerful RCTs, would be
considerably smaller, \eg~for $\rho = 0.6$ a reduction of
$13\%$ would be possible at most.
}

\section{Discussion}\label{sec:discussion}

In our work we studied the question, in what situations leveraging
prognostic information actually pays off in practice.  We presented a simple
and general setup in Section~\ref{sec:methods}, allowing us to assess the
theoretical properties of this adjustment method without making strong
distributional assumptions or limiting it to specific estimators.

In Section~\ref{sec:theoretic} we quantified the maximally attainable
benefit when adjusting for a prognostic score analytically, and contrasted
our findings with a more complex set-up in Section~\ref{sec:empeval}.  The
results suggest that leveraging prognostic baseline covariates reduces
residual variability, however the
magnitude of this reduction might often be irrelevant in practice.  These
situations can be characterised by small historical samples sizes (and as a
result smaller $\rho$) and/or small $R^2$ of the prognostic model on
historical data.

As a rough rule of thumb, sample size reductions of more than $20\%$ are
achievable with an $R^2 > .3$ on historical controls when there is a high
confidence in the prognostic score, with $\rho > .8$ say, requiring a large
number of historical controls and the absence of drift in $\PS$.  When there
is more uncertainty regarding the prognostic score, with $\rho \approx .6$
for example, an $R^2 > .5$ is necessary to obtain a $20\%$ reduction in
total sample size.
Likewise, the corresponding increase in precision of the treatment
effect estimate can be considered for fixed samples sizes.
It depends on the context whether or not such an increase
is relevant: It might be a game-changer in one setup but only marginally
interesting in other situations. 

While it is easy to estimate $R^2_\text{OOS}$ for historical controls, estimating
our model parameters $R^2$ and $\rho$ is less straightforward.  One possibility would be to perform an
interim analysis regressing the outcome $\rY$ on the prognostic score
$\hatPS$ on the trial controls $(\rY, \rX, \rz = 0)$ which, after
appropriate standardisation such that $\V(\hatPS) = 1$, gives an estimate
$\widehat{\pi \rho}$ for $\pi \rho$, which, together with an estimate of the
residual variance $\sigma^2$, can be plugged into (\ref{eqn:frac}).
In the absence of information about $\rho$, our interpretation of the
theoretical results presented here is that trial designers should
definitively look into the possibility of adjusting for an established
prognostic score when its $R^2_\text{OOS}$ has been demonstrated to exceed $.5$.

These findings are in agreement with earlier results quantifying the impact
of covariate adjustment on the necessary sample size in clinical trials. 
Adjusting for a single numeric covariate $X_1$ is a special case of our
model with $\pi \PS = \pi X_1$ and $\rho \equiv 1$, resulting in a ``design
factor'' of $1 - R^2$, meaning a sample size reduction to $(1 - R^2) \times
100\%$ of original sample size is possible
\citep{Borm_2007,Pocock_Assmann_Enos_2002,Cox_McCullagh_1982}. 
\added{
This ``design factor'' however disregards that the covariate
(or equivalentely the prognostic score) might be measured with error $\rho$ 
or that there might be potential distribution drift. 
}

%
Although accounting for prognostic information through adjustment for
$\hatPS$ seems rather unorthodox, a simpler version known as
post-stratification is well established.  For two strata, the prognostic
score $\hatPS \in \{0, 1\}$ is an indicator for the patient's stratum,
$\rho$ an unknown prognostic parameter, typically estimated from trial data. 
%
%
The rational is the same: Leveraging information from historical controls
(used to define reasonable strata) for reducing the residual variance while
safeguarding against distribution shift or incorrectly specified strata.
If available, such information further can be employed to randomise
patients into more homogeneous subgroups.

In the ANCOVA framework, an interesting practical question is when it will
be more beneficial to directly adjust for prognostic variables instead of
adjusting for a prognostic score, or even not to adjust at all
\citep{Lesaffre_Senn_2003}. We shall discuss this issue in more details.

Suppose we have $\n$ subjects in total and $k \geq 2$ prognostic covariates. 
(The lower bound is set at $2$ since the case $k = 1$ is without interest.)
The loss due to non-orthogonality, which we refer to as the \emph{imbalance
effect} is a random variable depending on the observed imbalance in the
trial.  However, choosing whether to fit the score or the covariates based
on an inspection of the data has the danger of increasing the type~1 error
rate.  Thus there is merit in making a pre-specified choice of model which,
in any case, is in line with ICHe9 recommendations.  It can be shown,
however, that the expected imbalance effect due to fitting $k$ covariates
compared to $1$ is $(\n-4) / (\n-3-k)$.  On the other hand the expected
inflation in the mean square error (MSE), which we refer to as the
\emph{mean square error effect}, due to fitting a score based on historical
data rather than the $k$ covariates on which it is based is $\sigma_1^2 /
\sigma_k^2 \geq 1$, where the numerator is the expected MSE for the
prognostic score and the denominator the corresponding MSE with all
covariates fitted.  Thus, by comparing the mean square error effect to the
expected imbalance effect one can make a decision.  Note that a third
factor is that the residual degrees of freedom for error will lead to the
$t$-table having to be entered at a less favourable point, the more
covariates are fitted.  As is discussed in an Appendix this further effect,
which we refer to as \emph{second order precision}, will favour the
prognostic score.

In summary, when the trial sample size $n$ is large and only a few prognostic variables
are studied, using ANCOVA without any involvement of historical data should
be preferred
\citep{Borm_2007,Pocock_Assmann_Enos_2002,Cox_McCullagh_1982}.
In situations where either the trial sample size $n$ is relatively small,
many and potentially unstructured prognostic variables shall be adjusted
for, and a large set of $\mathfrak{n}$ historical patient records is
available, it seems preferable to adjust for the prognostic score in
situations where $R^2 > .3$, because only one additional parameter needs to
be estimated in a classical statistical model.

An extension to non-normal models is not straightforward.  From a
computational point of view, the estimation of prognostic scores on
appropriate scales (log-odds or log-hazard ratios, for example), is possible
by application of some machine learning procedures, \eg~in model-based
boosting
\citep{Ridgeway_1999,Buehlmann_Hothorn_2007,Schmid_Hothorn_Maloney_2011}. 
Adjusting for such prognostic scores in logistic, proportional odds, or
proportional hazards regression models will lead to increasing power for
testing the null hypothesis $\trtparm = 0$ at the price of changing the
interpretation of the treatment effect estimate $\hat{\trtparm}$ from a
marginal to a conditional one
\citep{Robinson_Jewell_1991,Ford_Norrie_Ahmadi_1995,Ford_Norrie_2002,
Hernandez_Steyerberg_Habbema_2004,Rhian_2021}, owing to the fact that,
unlike in non-linear models, $\pi \PS$ can be absorbed into the error
term~(\ref{eqn:err}) in the linear model~(\ref{eqn:dgp}).

\section*{Acknowledgement}
Torsten Hothorn acknowledges funding from the Horizon~2020 Research and Innovation
Programme of the European Union under grant agreement number 681094, and is
supported by the Swiss State Secretariat for Education, Research and
Innovation (SERI) under contract number 15.0137.  The authors thank
Maria-Eleni Syleouni for initial simulation experiments testing the power of
prognostic score adjustment in her master thesis.

\vspace*{1pc}
\noindent{\bf{Author contributions}}
\noindent{\it
  Sandra Siegfried drafted the manuscript, contributed to the theoretical
  part, and performed empirical experiments.  Stephen Senn identified relevant
  earlier contributions and contributed the connection to ANCOVA provided in
  the appendix. Torsten Hothorn designed
  the study and developed the model.  All authors revised and approved the
  final version.
}
\vspace*{1pc}

\noindent{\bf{Supplementary material}}
\noindent{\it
  \proglang{R}~code to reproduce the empirical results is provided as
  supporting information.
  } %
\vspace*{1pc}
%

\noindent{\bf{Conflict of Interest}}
\noindent{\it
  The authors have declared no conflict of interest.
}
\vspace*{1pc}

\noindent{\bf{Data availability statement}}
\noindent{\it
Data sharing not applicable -- no new data generated.
}

\clearpage

\begin{appendix}

\section{Adjusting for covariates: Gains and losses}\label{sec:app}

An easy way to see the effect of fitting covariates on the efficiency of an
estimator is to consider adding a binary covariate (we shall take sex as an
example) to the analysis of a design that is currently balanced by treatment
with $2\N$ patients per arm, there being two arms in total.  If the covariate
is not fitted, the variance of the treatment contrast will be
\begin{eqnarray*}
\left ( \frac{1}{2\N} + \frac{1}{2\N} \right ) \sigma_0^2 = \frac{\sigma_0^2}{\N},
\end{eqnarray*}
where $\sigma_0^2$ is the within-treatment groups variance, which will be
estimated using $2\N - 2$ degrees of freedom and where the subscript 0 is
used to represent that no covariates have been fitted.  Now suppose that the
two sexes are equally well represented but having randomised and having
decoded the data, we see that the disposition of subjects by group and sex
is

{\centering
\begin{tabular}{lccc}\\
\hline
& Control & Treatment &\\
\hline
Females & $f$ & $2\N - f$ & $2\N$ \\
Males & $2\N - f$ & $f$ & $2\N$ \\
& $2\N$ & $2\N$ & $4\N = \n$ \\
\hline
\\
\end{tabular}
\\} \noindent
where the entries in the cells represents frequencies of patients of the
four types.  The within-sex stratum estimates now have variances
proportional to
\begin{eqnarray*}
\left( \frac{1}{f} + \frac{1}{2\N - f} \right) \sigma_1^2,
\end{eqnarray*}
where the subscript 1 is used to represent that one covariate has been
fitted. 
\added{Note, that one degree of freedom is lost if the fitting process uses
sex as a main effect in an analysis of covariance. However, strict stratification
estimates the variance within strata and loses one further degree of freedom.
Here we consider the former case, where the degrees of freedom available to
estimate this variance are now $2\N - 3$.} %
\deleted{
The degrees of freedom available to estimate this variance are now $2\N - 3$.
}
Clearly, the two within-stratum estimates are equally efficient
and should be weighted equally, that is to say by one half.  Thus the
combined estimate will have a variance equal to
\begin{eqnarray*}
\left( \frac{1}{2^2} \frac{1}{2^2} \right) \left( \frac{1}{f} + \frac{1}{2\N - f} \right) \sigma_1^2
&=& \frac{1}{2} \left( \frac{2\N}{f (2\N - f)}\right)\sigma_1^2\\
&=& \frac{\N}{f (2\N - f)}\sigma_1^2.
\end{eqnarray*}
Note that the divisor of this expression can be expressed as $\N^2 - (f -
\N)^2$ and that $-(f - \N)^2 \leq 0$,  so that the divisor reaches its maximum
when $f = \N$ that is to say the design is balanced, at which point the
variance will be $\sigma_1^2 / \N$.

We thus see that we can expect three consequences of fitting sex in the
model.  1) If sex is predictive we may expect $\sigma_1^2 < \sigma_0^2$. We
can refer to this as the \emph{mean square error effect}.  2) The variance
multiplier will be
\begin{eqnarray*}
\frac{\N}{f(2\N -f)} \geq \frac{1}{\N}
\end{eqnarray*}
with equality only being achieved in the case of perfect balance.  More
generally, we may expect some imbalance and so some loss in efficiency.  We
can refer to this as the \emph{imbalance effect}.  3) A completely predictable loss
is that the degrees of freedom associated with the relevant $t$-distribution
will be reduced by 1.  This, unlike the other two effects, is not an effect
on precision itself but an effect on our estimates of precision and may be
referred to as the \emph{second order precision effect}.  One way of judging it is
to compare the variances of the two $t$-distributions involved, using the fact
that in general this is $\nu / (\nu - 2)$ where $\nu$ is the degrees of
freedom.  In the case with no predictors,
we have $\nu = \n - 2$ and more
generally, if we have $k$ predictors, we have $\nu = \n - 2 - k$ so that the
general variance term is
\begin{eqnarray*}
\frac{\n - 2 - k}{\n - 4 - k},
\end{eqnarray*}
with this reducing to $(\n - 2) / (\n - 4)$ if $k = 0$, $(\n - 3) / (\n - 5)$
if $k = 1$.

More generally, for the cases where covariates may be continuous and there
may be more than one covariate but only two treatments, we may consider the
influence of these three factors in terms of the general variance estimator
$(\rX^\top\rX)^{-1}\sigma_k^2$.  Here $\rX_{\n \times (k + 2)}$ is the design
matrix for which we may assume, without loss of generality, that the first
column is an intercept carrier, the second is a treatment indicator and the
$k$ further columns, $k = 0, 1, 2, \dots$ are for the covariates.

This formulation includes not fitting covariates as a special case, for
which $k = 0$.  Note, however, that for the practical purpose of comparing
using a single score based on covariates to using the original covariates
themselves then the lowest value that is of any interest is $k = 2$.

The diagonal elements of the $(\rX^\top\rX)^{-1}$ matrix give the variance
multipliers and, given what we have said about the order of the columns, the
second of these is the multiplier for the variance of the treatment effect. 
We refer to this as $q_k$, where the subscript $k$ refers to the number of
covariates being fitted and not to the position in the matrix .  Thus the
variance of the treatment estimate is $q_k\sigma_k^2$.  Given $\n$ patients
it can be shown that we must have $q_k \geq 4/\n$.  For our previous example,
we had $\n = 4\N$ so we had $q_k \geq 1/\N$.

As covariates are added to the model and therefore columns are added to the
design matrix, the value of $q_k$ cannot reduce but may increase.  The
example with sex as a binary covariate illustrates this.  In a randomised
design the effect on $q$ is not predictable as the design matrix will vary
randomly but for normally distributed predictors the expected effect may be
described.  If the trial is balanced in the sense that there are the same
number of patients on each of the two arms but otherwise randomised, the
expected value is given by
\begin{eqnarray*}
\Ex(q_k) = \frac{4}{\n}\frac{\n -3}{\n - 3 - k}.
\end{eqnarray*}
Special cases are
\begin{eqnarray*}
&&\Ex(q_0) = \frac{4}{\n} \frac{\n - 3}{\n - 3} = \frac{4}{\n},\\
&&\Ex(q_1) = \frac{4}{\n} \frac{\n - 3}{\n - 4}.
\end{eqnarray*}
It thus follows that we have 
\begin{eqnarray*}
\frac{\Ex(q_k)}{\Ex(q_0)} = \frac{\n-3}{\n-3-k}, \\
\frac{\Ex(q_k)}{\Ex(q_1)} = \frac{\n-4}{\n-3-k},
\end{eqnarray*}
the second of these being relevant to the task of comparing adjustment for a
single score based on $k$ covariates to independently fitting them all them
all.  Note that this formula does not depend on the covariates being
generated by an independent process.  (The covariates, could, for example,
be correlated.) This is because, given $k$ predictors and assuming that the
set has no redundancy (the generating process is of rank $k$), they can be
replaced by $k$ orthogonal predictors which together will have the same
identical predictive value as the original $k$.  Furthermore, if we have a
predictive score, which is a linear combination of the predictors, then
given $k-1$ predictors and the score, the value of the remaining predictor
is completely determined and so redundant.  Thus the formula for
$\Ex(q_k)/\Ex(q_1)$ is valid for this case also.

Thus, consider making a decision as to whether to fit such a score.  A
relevant comparison is that of the ratio of the two expected mean square
errors to the ratio of the expected imbalance factors.  Thus  a sufficient
condition for fitting such a score would be
\begin{eqnarray*}
\frac{\sigma_1^2}{\sigma_k^2} \leq \frac{\n-4}{\n-3-k}, \ 2 \leq k \leq \n - 4
\end{eqnarray*}
where $\sigma_1^2$ is the mean square error fitting the score as a single
covariate.  Note that the right-hand side of the expression is an expectation but a
known quantity that must be greater than one (in expectation).  The left-hand side is a
random variable, which also ought to be greater than one, and some judgement
must be made by the modeller as to what it will be. 
The lower bound of $k$ is the lowest value of interest and the higher bound is
the highest for which the expression on the right-hand side is defined.

One could also try to incorporate the second order second order precision
effect into the decision process.  Note, however, that this is always in
favour of using the score rather than the  $k$ individual predictors. 
Therefore, if the condition above is satisfied it will definitely be an
advantage to fit the score.  This is why we refer to the condition as
sufficient.

However, it should be noted, that the expression provides a means of guiding
the choice between fitting $k$ predictors and fitting a linear combination
of them all.  If $k \geq 3$ it is possible that fitting a reduced set would
be better than either.

\end{appendix}

\end{document}